# Convection-induced compositional patterning at grain boundaries in irradiated alloys

G. F. Bouobda Moladje[1], R. S. Averback[1], P. Bellon[1], L. Thuinet[2]

[1]Department of Materials Science and Engineering,

University of Illinois at Urbana-Champaign, IL 61801

[2]Université de Lille, CNRS, INRAE, Centrale Lille,

UMR 8207 - UMET - Unité Matériaux et Transformations, F-59000 Lille, France

**Abstract**

We consider the stability of precipitates formed at grain boundaries (GBs) by radiation-induced segregation in dilute alloys subjected to irradiation. The effects of grain size and misorientation of symmetric-tilt GBs are quantified using phase field modeling. A novel regime is identified where, at long times, GBs are decorated by precipitate patterns that resist coarsening. Maps of the diffusional Péclet number indicate that arrested coarsening takes place when solute advection dominates over thermal diffusion right up to the precipitate/matrix interface, overwhelming capillary effects. This contrasts with liquid-solid mixtures where convection only accelerates coarsening.

Materials systems subjected to external forcing are often observed to self-organize into patterns [1-4]. Instabilities arise in these systems that can trigger the formation of transient structures, which then evolve into metastable or even stable steady-state patterns. Such patterns have been reported in solids and alloys subjected to irradiation [5,6] and to severe plastic deformation [7-9], resulting in microstructures with emergent pattern length scales and symmetry. Notable examples of *defect* patterns are void, bubble and walls of dislocations, which derive from anisotropic diffusion of point defects and defect clusters in irradiated crystals [10,11], and nanoscale *compositional* patterns in alloys, which are due to the non-local character of forced chemical mixing during irradiation or plastic deformation[12]. Convection can also trigger patterning, but this is traditionally studied in fluids, where hydrodynamic instabilities are often present [4]. In particular, random and turbulent [13-16] or chiral [17] shear flows in binary fluid mixtures promote domain stretching and break-up, resulting in arrested coarsening and steady-state patterning for large enough ratios of chemical convection to chemical diffusion, i.e., at high enough chemical Péclet number, *Pe*. In the case of solid particles in a fluid matrix, however, both experiments [18] and theory [19,20] indicate that convection only accelerates coarsening, even for a high *Pe* number. This latter observation can be rationalized by noting that solid particles in a fluid flow do not stretch or break-up, while convection promotes solute transport and particle coagulation. The role of chemical convection on the coarsening of solid particles in solid matrices has received much less attention, despite its relevance to various materials processes: the transport of ionic species in battery electrodes and electrolytes during charge and discharge cycles [21]; the chemical mixing forced by plastic deformation in crystalline alloys [7,22,23]; the transport of mass to surfaces in energetic displacement cascades [24-26]; and the flow of point defects to sinks under irradiation [4]. We thus consider here the question of whether convection in solid-solid systems can also induce patterning. Using a simple model for point defect and chemical transport in an irradiated alloy, along with phase field simulations, we report on a novel compositional patterning phenomenon at grain boundaries (GBs) and show that it results from solute advection to GBs coupled with anisotropic solute diffusion at GBs. More specifically, we illustrate that arrested coarsening is a consequence of a high diffusional Péclet number along the GBs, which suppresses the effects of equilibrium capillary forces.

The generic microstructure investigated in this work consists of a flat grain boundary formed by two abutting grains in a dilute A-B alloy. Irradiation of that alloy produces isolated point defects homogeneously. These point defects, vacancies and interstitials can then recombine or diffuse to the GB, which is treated as a perfect sink (see Section SI in Supplemental Material). Continuous irradiation therefore sets in permanent point defect fluxes to the GB. For the alloy considered here, such point defect migration leads to solute transport to sinks through flux coupling, also known as radiation-induced segregation [27-29]. This solute advection can trigger solute precipitation at the sinks. The central question of the present work is to determine the resulting precipitation microstructure and whether it can undergo self-organization under appropriate irradiation conditions. Phase field modeling is used for that purpose, since it can include the main physical processes relevant to irradiation and evolve systems of sufficient sizes over long timescales [30]. The state of the system is described by four phase field variables, $X_V$, $X_I$, $X_A$ and $X_B$, representing the molar fraction of vacancies, interstitials, solvent and solute atoms, respectively. The kinetic evolution equations for these field variables are given by a kinetic phase-field model that includes radiation effects and flux coupling [31]:

$$\frac{\partial X_d}{\partial t} = \nabla . \sum_\alpha \sum_\beta \frac{l_{\alpha\beta}^d X_d}{k_B T} \left( sign(d) \nabla \frac{\delta F}{\delta X_\beta} + \nabla \frac{\delta F}{\delta X_d} \right) + K_0 - K_R - K_s^d$$
$$\frac{\partial X_\alpha}{\partial t} = \nabla . \sum_d \sum_\beta \frac{l_{\alpha\beta}^d X_d}{k_B T} \left( \nabla \frac{\delta F}{\delta X_\beta} + sign(d) \nabla \frac{\delta F}{\delta X_d} \right) + \zeta_{noise}$$
(1)

where $\alpha, \beta = A, B$, and $d = V, I$ with $sign(V) = -1$ and $sign(I) = 1$. $K_0$ represents the point defect production rate in displacements per atom (dpa) per second, T the temperature, $K_R$ the mutual recombination rate between vacancies and interstitials, and $K_s^d$ the point defect annihilation rate at the sink. $\zeta_{noise}$ is a Gaussian random noise function simulating thermal fluctuations to suppress trapping in shallow metastable configurations. $\left( l_{\alpha\beta}^d \right)$ represents the matrix of Onsager transport coefficients; it is symmetric, positive and depends on the alloy temperature and composition. The total free energy of the system $F$ includes local chemical interactions, and long-range elastic interactions. The chemical free energy is written as the sum of the free energy of the homogeneous alloy, using a regular solid solution model, and a gradient energy term accounting for heterogeneities in the concentration field of chemical species [31]. The elastic energy results from the interactions between point defects and sinks, and is calculated via microelasticity theory [32] by considering an elastically homogeneous medium. The strain energy due to the lattice misfit is neglected here for the sake of simplicity. The atomic mixing forced by nuclear collisions induced by irradiation, i.e., ballistic mixing [33], is deliberately left out, so that pattern evolutions detailed below cannot be associated with such mixing. This is of note since previous modeling studies have indeed demonstrated that ballistic mixing can trigger compositional patterning in irradiated alloys [34-37]. For computational efficiency, the phase field equations for only $B, V, I$ in Eq. (1) are solved, in 2D cartesian coordinates ($x_1$,$x_2$), using a forward Euler method with an adaptive timestep [31] (see Fig. S1 in Supplemental Material [38]).

Parameters for the model A-B immiscible alloy system investigated here mimic a dilute Al-Sb alloy, but mainly they were selected to promote strong flux coupling (see Table S1 [38]). Specifically, the large solute-vacancy binding energy $E_{Sb-V}^b = 0.3 \; eV$ [39] results in a strong advection of the solute to the GB by vacancy fluxes at 300 K. The drag ratios of Onsager coefficients calculated by the transport coefficient code KineCluE [40] are $l_{VB}^V/l_{BB}^V \approx 1$ and $l_{VB}^V/l_{VV}^V \approx 1$ (see Fig. S2). In contrast, interstitials do not promote Sb segregation since oversized solutes such as Sb in Al do not form mixed dumbbells; we thus set $l_{BB}^I = l_{AB}^I = 0$ in Eq. (1). The vacancy-driven solute-drag flux (the second term in the double summation in Eq. 1(c)) is opposed by a back diffusional flux (the first term in the double summation) once the solute depletion in the matrix drops below the equilibrium solubility limit, $X_B^{eq} = 1.75 \times 10^{-4}$ at T = 300 K. Note that the diffusion coefficient controlling this back flux is enhanced by the point defect supersaturation created by irradiation, and it is referred to here as $D_B^{RED}$. In addition to these fluxes, which are primarily perpendicular to the GB, solute is also mobile in the grain boundary.

Two different models for the GB were investigated. The first model corresponds to symmetric tilt grain boundaries (STGBs), modeled as an array of equally spaced edge dislocations. The misorientation $\theta$ of an STGB is related to the spacing between dislocation cores, $h$, as $h = b/2\sin(\theta/2)$, $b$ being the module of the Burgers vector. Misorientations from 2.4° ($h = 24b$) to 14.3° ($h = 4b$) were studied. Point defect concentrations within the GB capture region were held equal to bulk equilibrium values, $X_V^{eq} = 1.5 \times 10^{-9}$ and $X_I^{eq} \approx 0$, respectively. The capture zone was set to $4b$, corresponding to the dislocation core width. The values of $X_V^{eq}$ and $X_I^{eq}$ are modified by the elastic strain field around the dislocation cores, see Fig. S3 [38], but this effect is relatively small. The solute diffusion coefficient along the GB, $D_B^{GB}$, is therefore approximately equal to $D_B^{b,eq}$, the equilibrium bulk solute diffusion coefficient. The second type of GB used in the simulations mimics large angle GBs and relies on a continuum description, where the sink absorption region is a straight stripe of width $2b$.

Table 1 summarizes the microstructures reached at long times, here corresponding to an irradiation damage of 10 dpa, for various grain sizes. For the STGBs, solute concentration builds up around each dislocation core, leading eventually to solute-rich precipitates decorating each core. Two "trivial" structures can thus be expected at long times by comparing the precipitate diameter, $d_p$, to $h$: When $d_p < h$, e.g., $d = 13.44$ nm and $\theta = 2.38^o$ in Table 1, the precipitates do not overlap, resulting in a structure, hereafter referred to as Type 1, where each dislocation core is decorated by one solute-rich precipitate; in contrast, when $d_p > h$, individual GB precipitates overlap and merge, eventually forming a continuous wetting layer along the GB, labeled as structure of Type 2, e.g., $d \geq 80.64 nm$ and $\theta = 14.32^o$ in Table 1 (see also Fig. S4).

More interesting structures are observed for intermediate grain sizes and misorientations.

These structures are comprised of a finite number of precipitates, larger than 1 but smaller than the number of dislocation cores, see for instance $d \leq 80.64$ nm for $\theta = 9.55°$ in Table 1. These GB structures, referred to as Type 3 structures, developed differently for small and large misorientations. For small misorientations, Type 3 structures began as Type 1 structures that underwent partial coarsening with time, see Fig. 1, but coarsening became arrested at longer times. For larger misorientations, the Type 3 structures were first continuous GB solute-rich films at low doses (<0.1 dpa), i.e., similar to Type 2 structures, but precipitates subsequently formed by decomposition along the GB. The size of these precipitates, however, stabilized at long times, see Fig. 2. For continuous sinks, see last column in Table 1, Type 3 structures are again observed, following a mechanism very similar to that observed for large tilt boundary misorientations, i.e, spinodal-like decomposition along the GB and arrested coarsening. Decomposition occurs at larger doses (> 1dpa) for these boundaries, in comparison to STGB's, since it is not assisted by heterogeneities in this case, see Fig. S5. Finite size effects and periodic boundary conditions were found to affect the exact linear density of precipitates found in Type 3 structures, as expected (see Fig. S6) but not their main features nor their unexpected resistance to coarsening. We identify Type 3 structures as GB compositional patterns since these structures are highly stable and the separation distance between precipitates is an emergent length scale, distinct from the microscopic scales $b$ and $h$, and the macroscopic scales set by the system sizes along $x_1$ and $x_2$.

The stability of selected Type 3 structures was confirmed by increasing the irradiation dose to 30 dpa. For a structure with two precipitates for $\theta = 9.55°$ and $d$=13.4 nm, the total solute content in each precipitate remains largely constant up to 20 dpa, see Fig. S7. Some very slow coarsening may occur beyond that dose, although much longer simulations would be required to confirm it. The stability of the Type 3 structures was further examined by comparing coarsening kinetics in this system with that in a system without convection, i.e., by switching off irradiation at some finite dose and subsequently annealing the Type 3 structures thermally using Eq. (1) without sinks. In a first annealing scheme, vacancy concentrations were allowed to relax, resulting in nearly isotropic solute diffusion. In the second scheme, vacancies were frozen-in, thus retaining the anisotropic diffusion present during irradiation. For both annealing schemes, coarsening was soon observed, see two examples for discrete and continuous GBs in Table S2. The characteristic times for coarsening were accelerated in scheme 1 annealing by factors ≈ 1,300 and ≈ 3,000, for the above two GBs, respectively, relative to the irradiated system. These acceleration factors remain large, ≈ 200 and ≈ 150, even with scheme 2 annealing. It is thus concluded that diffusion anisotropy influences the stability of Type 3 structures, but it is not the dominant factor. More important, as we now discuss, is the convective flux arising from strong coupling between solutes and vacancies.

To understand these various results, we begin by considering the standard description of solid particle coarsening in fluids in the presence convection [19,20]. The diffusional Péclet number in our system is defined as $Pe = (vd_{adv})/D_B^{RED}$ where $v$ is the solute drag velocity, and $d_{adv}$ the characteristic distance over which advection takes place. $Pe$ is neither uniform in space nor independent of time since solute segregation affects vacancy fluxes and vacancy concentrations, therefore affecting both convection and diffusion. At early times, however, before significant solute redistribution takes place, an analytical expression for a system-averaged Péclet number was derived, $\langle Pe \rangle_o \approx (X_B K_0 d^2)/(\phi l_{VB}^V X_V^{irr})$, where $\phi$ is the thermodynamic factor of the alloy, see Section SIII [38]. Parametric phase field simulations confirm the scaling of $\langle Pe \rangle_o$ with $(K_0 d^2)$, thus rationalizing the large effect that the grain size $d$ has on stabilizing Type 2 structures. For the present parameters, the initial $Pe$ number is very high, in excess of $10^4$. Once solute redistribution takes place, the phase field simulation results are used to map out $Pe(x, t)$. From the definition of the diffusional Péclet number and Eq. (1), one obtains $Pe(x, t) = |\nabla(\delta F/\delta X_V)|/|\nabla(\delta F/\delta X_B)|$. Far from the GB, $Pe \approx 1$, as solute advection by RIS is balanced by back thermal diffusion. Close to the precipitate/matrix interfaces, diffusion is expected to dominate over convection, resulting in $Pe < 1$ or $\ll 1$ [20]. In Type 3 structures, however, $Pe$ reaches large values between precipitates near dislocation cores, see Fig. 3(b), since these sinks draw large advection currents, and since $D_B^{GB}$ is small recall, $D_B^{GB} \approx D_B^{eq}$. Moreover, maps of solute concentrations, e.g., Fig. 3(a,c), reveal remarkable features along the GB: (i) The solute concentration far exceeds the bulk solubility limit, here by up to 2 orders of magnitude; and (ii) no significant capillary effects can be detected, i.e., the solute profile just outside the precipitate/matrix interface does not show any dependence with the precipitate size.

The above results lead us to propose that the high value of $Pe$ along the GB, between precipitates, is the determinant factor for the stabilization of Type 3 structures. When this $Pe$ value is large, the width of the boundary layer separating the diffusion-dominated region, where $Pe < 1$, from the convection-dominated region, where $Pe > 1$, becomes smaller than the half-width of the precipitate-matrix interface, see Fig. 3(c). Consequently, convection-induced solute segregation overwhelms equilibrium capillary effects and suppresses coarsening. This hypothesis was tested by artificially increasing the equilibrium vacancy concentration to increase GB diffusivity while reducing convection. Coarsening of Type 3 structures was clearly observed in those cases. Upon arbitrarily increasing $X_V^{eq}$ from $1.5 \times 10^{-9}$ to $10^{-6}$, for instance, an additional low dose of 1.1 dpa led to the full dissolution of one of the three precipitates shown in Fig. S5 for the continuous GB sink, in contrast to the absence of coarsening for doses up to at least 20 dpa for the correct $X_V^{eq}$.

The present 2D simulation results provide a framework to anticipate the Type 3 structures that could form in 3D systems. At low *Pe* number, a Rayleigh-Plateau instability, which does not take place in 2D systems [41], could destabilize continuous tubular precipitates along GB misfit dislocations, resulting in individual precipitates decorating GB. This instability should be promoted by the fact that the relevant solute diffusion coefficient would be the one along the dislocation cores, thus larger than those between dislocation cores considered here. Single-component Fickian diffusion to a linear sink in cylindrical geometry, however, produces steady-state fluxes that scale as 1/r, in contrast to constant fluxes for 2D cartesian geometry. For low GB misorientations, larger RIS solute advection should thus promote the stabilization of tubular precipitate structures. Testing these predictions is left for future work.

While the present results do not include ballistic mixing, they should approximate irradiation situations employing light ions, such as protons or energetic (MeV) electrons, as atomic transport occurs arises primarily from defect mobility and not from recoil collisions. To the best of our knowledge there is no direct experimental evidence that irradiation could induce precipitate patterning at GBs. It is worth noting that solute precipitate structures have recently been observed at misfit dislocations and interfaces [42,43]. These structures, however, were rationalized as resulting from equilibrium segregation followed by spinodal decomposition [42] or from equilibrium wetting at misfit dislocation intersections [43], in contrast to the nonequilibrium kinetic stabilization identified here. The present work suggests that different GBs will respond differently since their distinct structures will affect their defect sink properties and diffusion coefficient $D_B^{GB}$. It is also envisioned that other nonequilibrium systems with internal sinks are susceptible to similar self-organization reactions in the presence of convective flows, e.g., battery electrodes during charging or discharging cycles, or alloys subjected to severe plastic deformation.

**Acknowledgements**

The research was supported by the U.S. Department of Energy, Office of Science, Basic Energy Sciences, under Award No SC0019875. This work made use of the Illinois Campus Cluster, a computing resource that is operated by the Illinois Campus Cluster Program (ICCP) in conjunction with the National Center for Supercomputing Applications (NCSA) and which is supported by funds from the University of Illinois at Urbana-Champaign. Stimulating discussions with Drs. Charles Schroder and Peter Voorhees are gratefully acknowledged.

**Contributions**: All authors contributed significantly to the work. The primary contributions are as

**Figures and Table**

| d (nm) | Misorientation angle $\theta = b/h$ | | | | |
| --- | --- | --- | --- | --- | --- |
| | 2.38° | 4.77° | 9.55° | 14.32° | continuous sink |
| 13.44 nm | ◆ | ● | ● | ● | ○ |
| 26.88 nm | ◆ | ● | ● | ● | ☆ |
| 40.32 nm | ● | ● | ● | ● | ■ |
| 53.76 nm | ● | ● | ● | ✱ | ■ |
| 80.64 nm | ● | ● | ● | ■ | ■ |
| 94.08 nm | ● | ● | ■ | ■ | ■ |
| 107.52 nm | ● | ■ | ■ | ■ | ■ |

■: solid film; ✱: single precipitate; ●, ○: 2 precipitates; ☆: 3 precipitates; ◆: 4 precipitates

Table 1: GB precipitate structures reached after an irradiation dose of 10 dpa, as a function of the grain size and the GB structure.

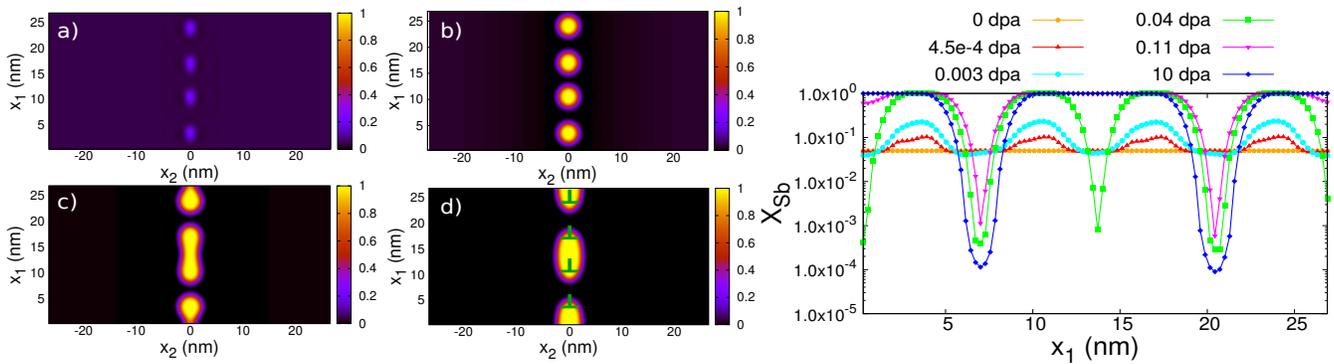

Figure 1: Evolution of the Sb atomic fraction map for $\theta = 2.38°$ and $d = 53.76$ nm for a) 0.003 dpa, b) 0.04 dpa, c) 0.11 dpa, d) 10 dpa, and the corresponding profiles along the GB.

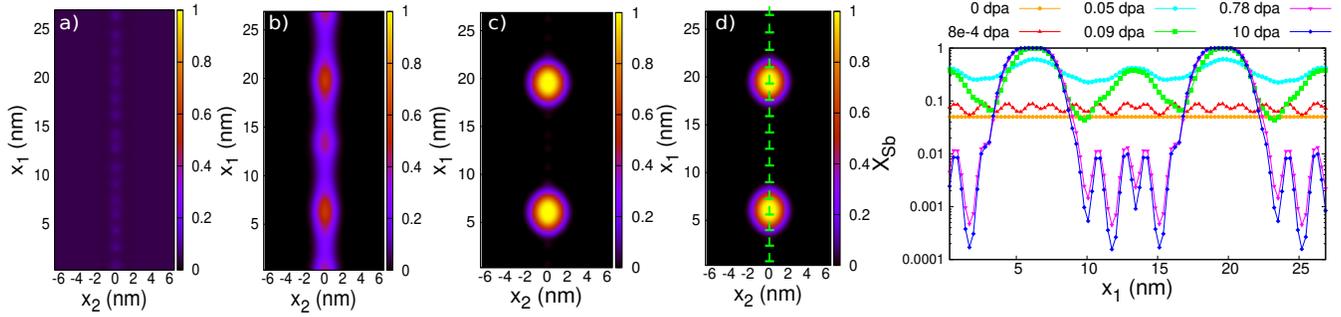

Figure 2: Evolution of the Sb atomic fraction map for $\theta = 9.55°$ and $d = 13.44$ nm for a) $8 \times 10^{-4}$ dpa, b) 0.05 dpa, c) 0.78 dpa, d) 10 dpa, and the corresponding profiles along the GB.

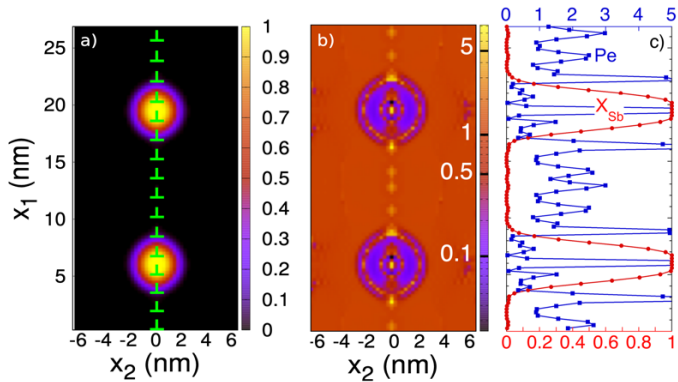

Figure 3: Maps of a) solute atomic fraction $X_{Sb}$ and b) Péclet number $P_e$ for STGB with $\theta = 9.55°$ and $d = 13.44\ nm$ at 10 dpa. c) Profiles of $X_{Sb}$ (lower horizontal axis, red color online) and $Pe$ (upper horizontal axis, blue color online) along $x_1$ on the GB ($x_2 = 0$), truncated at $Pe \leq 5$ for clarity; the high value of $Pe$ at the center of precipitate is due to the vanishingly small diffusion imposed by symmetry. The $x_1$ vertical axis is the same for a), b), and c).

**Supplemental Material:**

**Convection-induced compositional patterning at grain boundaries in irradiated alloys**

G. F. Bouobda Moladje[1], R. S. Averback[1], P. Bellon[1], L. Thuinet[2]

[1]Department of Materials Science and Engineering,

University of Illinois at Urbana-Champaign, IL 61801

[2]Université de Lille, CNRS, INRAE, Centrale Lille,

UMR 8207 - UMET - Unité Matériaux et Transformations, F-59000 Lille, France

**SI: Simulations setup and model parameters**

The necessary parameters for the phase field simulations are described in this section and given in Table S1. The elastic constants and point defects (PDs) relaxation volumes used for the computation of the elastic interactions between PD and dislocations correspond to that of the pure Al matrix and were taken from [S1]. The transport coefficients were calculated in the limit of dilute systems using the KineCluE code [S2] and, for simplicity, they were used for the nominal concentration of 5 $at.\%$ in the present generic study. When quantitative modeling for specific alloy is the goal of the study, transport coefficients for concentrated alloys can be obtained from either experiments or atomistic simulations. The evolution of the ratios of the transport coefficients $l_{VB}/l_{VV}$, $l_{VB}/l_{BB}$ as a function of the temperature is represented on Figure S2. The ordering energy $\Omega_{mix}$ is related to the solubility limit $X_B^e$ in the case of a regular solid solution by $k_B T X_B^e + \Omega_{mix}(1 - X_B^e)^2 = 0$. $\Omega_{mix}$ was then deduced from this last expression using the value of $X_B^e = 0.11$ at the eutectic point of the Al-Sb phase diagram at 931K [S3]. The thermal equilibrium concentration of self-interstitial atoms was calculated in pure Al using $X_I^{eq} = exp(-E_f^I/k_B T)$ where $E_f^I$ is the interstitial formation energy. In the case of vacancies, due to the formation of V-B complex, we used the expression given by Eq. 6 in [S4], which allows to take into account the change of the defect concentration in the presence of solute. The gradient energy coefficient $\kappa$ was calculated using its relationship to the ordering energy [S5]: $\kappa = \Omega\, r_1^2/3$, where $r_1$ is the distance between nearest neighbor atoms, corresponding here to the Burgers vector $b$. This coefficient allows to control the interface thickness and was artificially increased in our simulations in order to have a sufficiently diffuse interface, as usually done in phase field studies to suppress numerical instabilities [S6-S8]. The PD atomic fractions were maintained at their equilibrium values inside the sinks, dislocation cores and continuous sink, to reproduce the behavior of a perfect sink. Unless stated otherwise, the starting condition is a uniform solid solution. The solvent concentration in the grain interior is deduced from the lattice site conservation equation $X_A + X_B + X_V - X_I = 1$, while within sinks it is approximated by $X_A \approx 1 - X_B$. Selected simulations performed without this latter approximation show that it only shifts the initial kinetic evolution of solute segregation and precipitation at sinks by a small dose, but it does not affect the long-term microstructures, see Fig. S1.

| | |
|---|---|
| Temperature $T$ [K] | 300 |
| Nominal composition $X_B^{nom}$ [at.%] | 5% |
| Transport coefficients $l_{AA}^V, l_{AB}^V, \frac{l_{BB}^V}{X_B}, l_{AA}^I, l_{AB}^I, l_{BB}^I$ [$m^2/s$] | $1.2 \times 10^{-16}, \approx 0, 2.6 \times 10^{-12}, 8 \times 10^{-9}, 0, 0$ |
| Point defect generation rate $K_0$ [dpa/s] | $10^{-3}$ |
| Module of the Burgers vector $b$ [nm] | 0.28 |
| Elastic constants $C_{11}, C_{12}, C_{44}$ [GPa] | 106.5, 60.4, 27.8 [S1] |
| Relaxation volume of point defects $\omega_V, \omega_I$ | $-0.4 V_{at}, 2.35 V_{at}$ [S1] |
| Atomic volume $V_{at}$ [$m^3$] | $1.66 \times 10^{-27}$ |
| Ordering energy $\Omega_{mix}$ [eV] <br> Coefficient of the gradient energy $\kappa$ [$eV \cdot nm^2$] | 0.223 <br> 0.37 |
| Formation energy of point defects $E_f^V, E_f^I$ [eV] | 0.67, 1.57 [S9] |
| Equilibrium fraction of point defects $X_V^{eq}, X_I^{eq}$ [at.%] | $1.5 \times 10^{-9}, 2.98 \times 10^{-27}$ |
| Solubility limit $X_B^e$ [at%] | $1.75 \times 10^{-4}$ |
| Spinodal limits $X_B^{spinodal}$ [at%] | $[6.16\% - 93.84\%]$ |

Table S1: Input parameters for the PF simulations

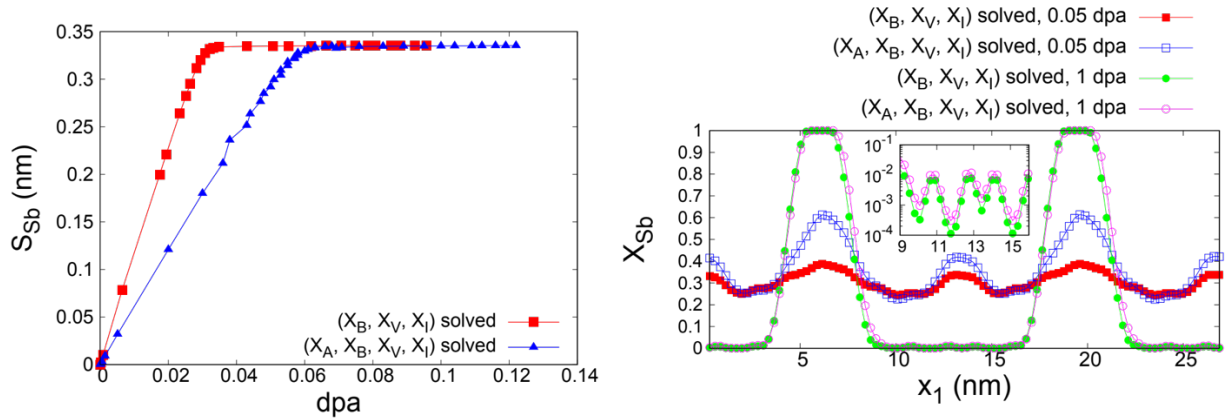

Figure S1: a) Evolution with the irradiation dose of the total solute segregation length $S_{Sb}$ as defined in [S10] for two type of simulations. Type 1: the evolution equations of the variables $(X_B, X_V, X_I)$ are solved while the variable $X_A$ is deduces from the lattice site conservation equation $X_A + X_B + X_V - X_I = 1$ in the bulk, and inside the sink it is approximated by $X_A + X_B \approx 1$ (red curve) ; Type 2: the evolution equations of the order parameters $(X_A, X_B, X_V, X_I)$ are all solved (blue curve). b) Solute concentration profile along the GB for $\theta = 9.55°$ as a function of the dose for the simulation type 1 and 2. The kinetics of solute segregation is slowed down by about 0.02 dpa in simulation type 2 compared to that of simulation type 1 as illustrated by the $S_{Sb}$ evolution. There are however no significant changes in the segregation profiles at a dose of 1 dpa, and beyond.

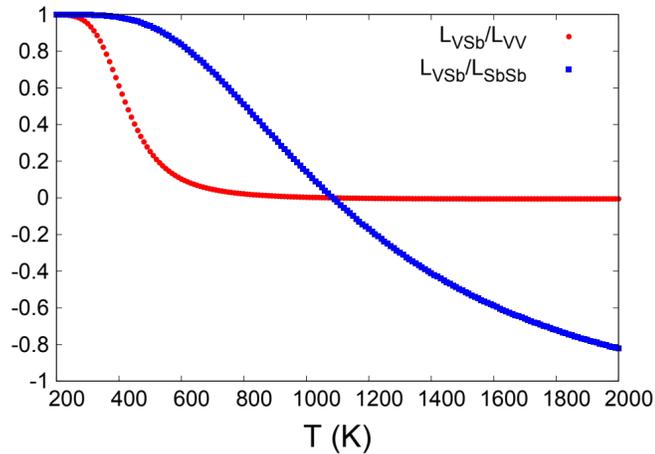

Figure S2: Evolution of the drag ratios of the Onsager transport coefficients as a function of the temperature computed using the KineCluE code [S2]. The ratios $l_{VB}/l_{VV}$, $l_{VB}/l_{BB}$ are positive when the B atoms are dragged by vacancies.

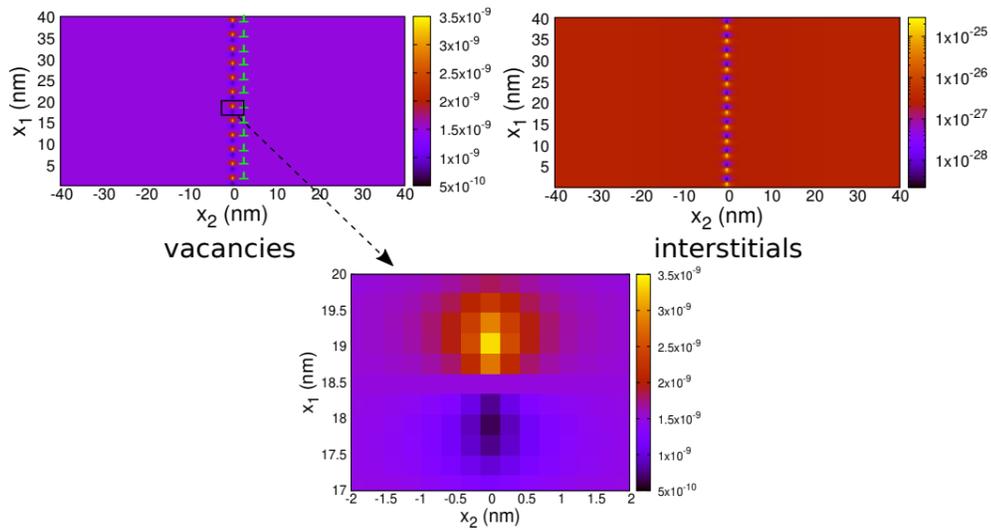

Figure S3: Atomic fraction maps of vacancies and self-interstitial atoms (SIAs) along the GB at the thermal equilibrium. There is enrichment (respectively depletion) of vacancies in the compression (respectively tension) region of each dislocation while it is the opposite in the case of SIAs, which is expected when elastic interactions are taken into account.

**SII. Precipitate structures at grain boundary under irradiation**

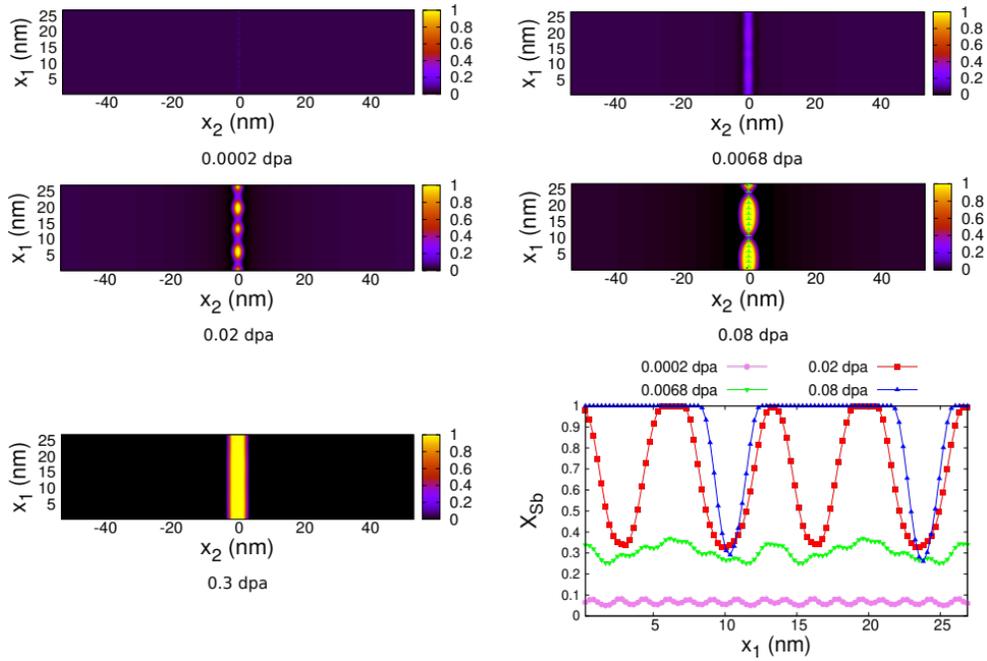

Figure S4: Evolution of the atomic fraction map of the solute and the corresponding profile along the GB as a function of the dose for $\theta = 9.55°$ and $d = 107.52 nm$, showing the kinetics of the formation of a continuous wetting layer along the GB. At very low doses, the solute is segregated at each dislocation, then the segregated regions overlap and form a continuous GB solute-rich film. It undergoes a decomposition along the GB analogous to spinodal decomposition assisted by the dislocation cores. The precipitates then grow and merge until the formation of the continuous wetting layer.

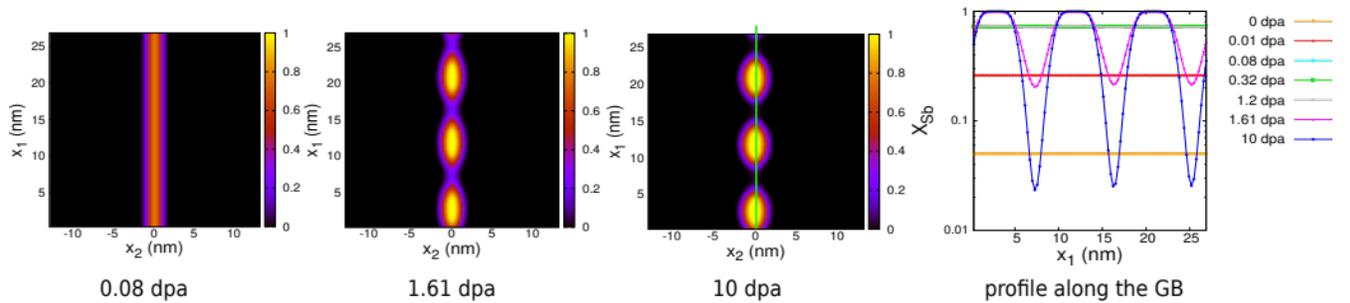

Figure S5: Evolution of the Sb atomic fraction map as a function of the dose and the corresponding profiles along the continuous GB, $d = 26.88 nm$.

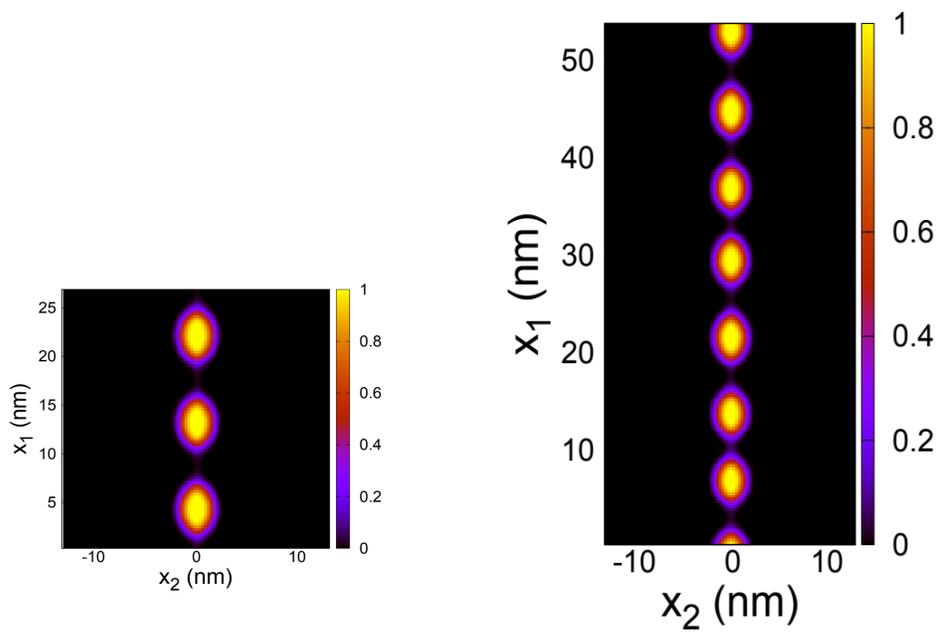

Figure S6: Effect of the simulation box size along the $x_1$ dimension on the number of precipitates formed along the continuous sink at 10 dpa.

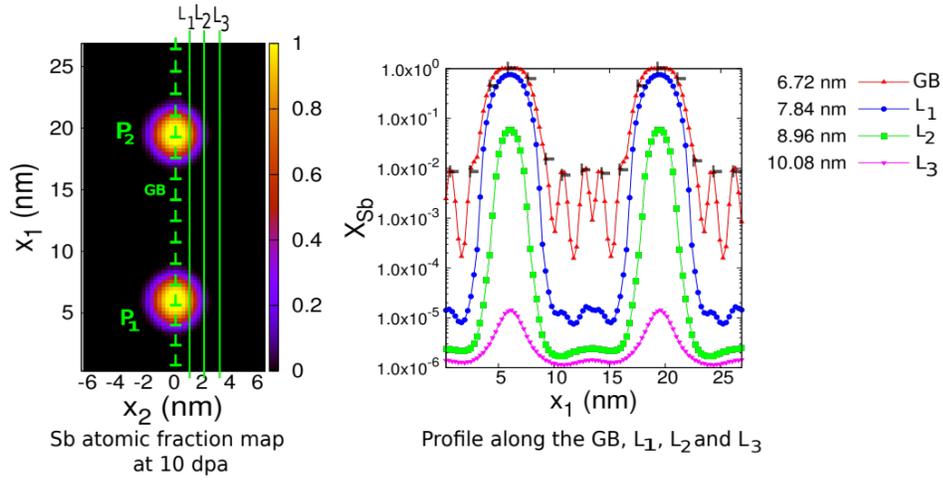

(a)

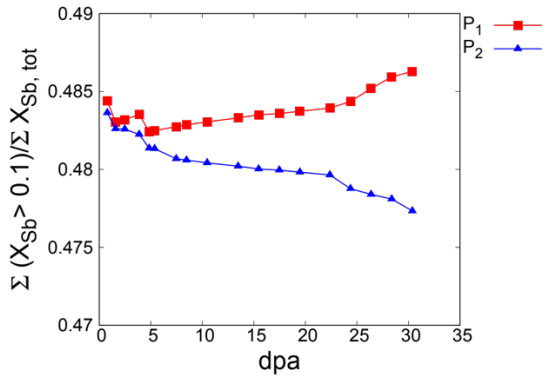

(b)

Figure S7: a) Atomic fraction map of the solute after reaching a dose of 10 dpa and the corresponding profiles along the GB and adjacent lines L1, L2, L3 for $\theta = 9.55^o$ and $d = 13.44\ nm$; b) evolution of the fraction of solute contained in each precipitate as a function of the dose. The precipitate interior was defined such as $X_{Sb} > 0.1$.

|  | Characteristic time to observe a change of the precipitate size of 1% starting from a system irradiated to 5 dpa for $\theta = 9.55^o$ and $d = 13.4$ nm | Characteristic time to observe a change of the precipitate size of 1% starting from a system irradiated to 2 dpa in the case of the continuous GB for $d = 26.8$ nm |
| --- | --- | --- |
| Irradiation | 20.4 dpa | 3.16 dpa |
| Annealing scheme 1 | 0.015 dpa | 0.0011 dpa |
| Annealing scheme 2 | 0.092 dpa | 0.023 dpa |

Table S2: Characteristic doses (or time) for coarsening kinetics during irradiation and thermal annealing. The stability of the Type 3 structures was examined by comparing coarsening kinetics in this system with that in a system without convection, i.e., by switching off irradiation at some finite dose and subsequently annealing the Type 3 structures thermally. In a first annealing scheme, the system-average vacancy concentration was kept but the vacancies were free to redistribute themselves, i.e., one solved Eq. (1) with $K_0 = 0$ and $X_I = 0$ everywhere and without sinks. In addition, the B-V binding was set to zero, and thus the vacancy concentration became quickly uniform, resulting in largely isotropic diffusion. In scheme 2, we retained the diffusion anisotropy during annealing by solving Eq. (1a) but freezing the vacancy concentration field at its value when irradiation was switched off. For both annealing schemes, coarsening was soon observed in the simulations, as illustrated by two examples for $\theta = 9.55^o$, $d = 13.4$ nm, and in the case of the continuous GB for $d = 26.8$ nm. For thermal annealing cases, an equivalent irradiation dose was defined using the displacement rate value employed for the irradiation $K_0 = 10^{-3}\ dpa/s$.

**SIII: Derivation of the Péclet number at small dose**

The diffusional (or solute) Péclet number of the system is defined as:

$$Pe = (v d_{adv})/D_B^{RED} \tag{S1}$$

where $v$ is the solute drag velocity and $d_{adv}$ the characteristic distance over which advection takes place. Using standard diffusion analysis for a dilute irradiated alloy [S11], one can express the drag velocity in terms of the vacancy gradient:

$$v = (l_{VB}^V |\nabla X_V^{irr}|)/X_V^{irr} \tag{S2}$$

where $|\nabla X_V^{irr}|/X_V^{irr}$ is evaluated at or close to the GB, and we used $d_{adv} = d$, which holds for small GB separation distances $d$. In the present work, the expression for $Pe$ can be further simplified since the dominant contribution to vacancy diffusion is that of bound V-B complexes:

$$D_B^{RED} = \phi X_V^{irr} l_{VB}^V / X_B \tag{S3}$$

where $\phi$ is the thermodynamic factor, and thus:

$$Pe = (X_B |\nabla X_V^{irr}| d)/\left(\phi (X_V^{irr})^2\right) \tag{S4}$$

Moreover, for small grain sizes, a simple analytical expression for the vacancy gradient in a homogeneous alloy in the absence of recombination is given by [S12,S13]:

$$|\nabla X_V^{irr}| \propto (K_0 d)/D_V \tag{S5}$$

where the average vacancy diffusion coefficient is $D_V = l_{VB}^V / X_V^{irr}$ [S4]. The linear scaling of $|\nabla X_V^{irr}|$ with $K_0$ and $d$ has been confirmed in the present phase field simulations, see Fig. S8. We finally obtain:

$$Pe \approx (X_B K_0 d^2)/\left(\phi l_{VB}^V X_V^{irr}\right) \tag{S6}$$

For all parameter values considered in this study, we find that $Pe \geqslant 10^4$ at that initial quasi-steady state, i.e., before any solute redistribution has taken place. The initial solute evolution in the bulk is thus strongly dominated by its convection via radiation-induced segregation to the GB. The initial rate of solute segregation on the sink, defined as in [S10], was found in the PF simulations to increase linearly with $d$ for small $d$ values, see fig. S9, in agreement with the above derivation indicating that $|\nabla X_V| \propto d$. At longer irradiation times, back diffusion progressively opposes solute drag, leading to steady states where $Pe \approx 1$ away from the GB.

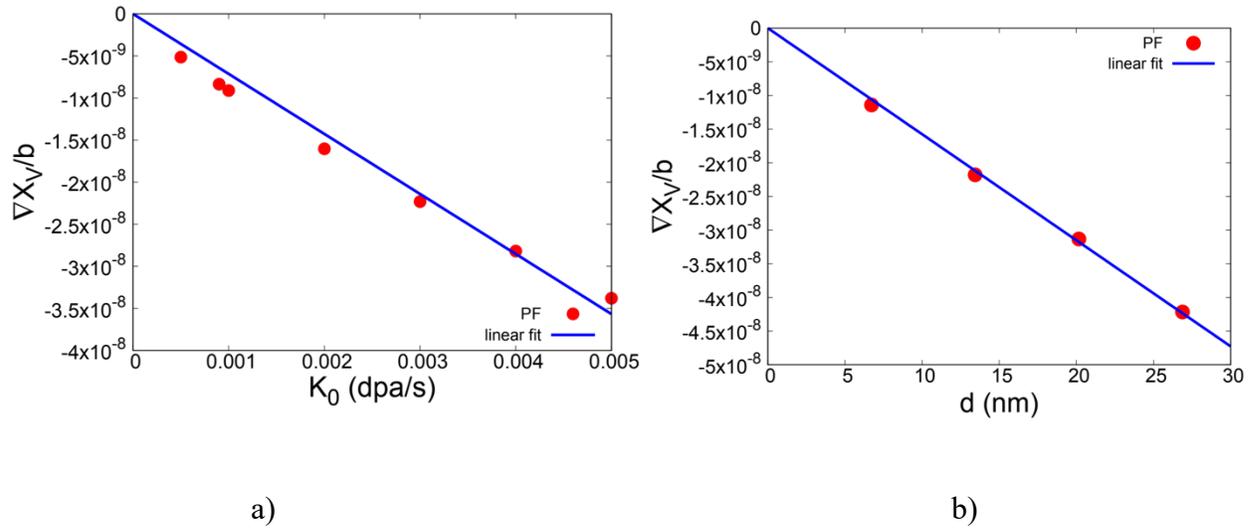

a)            b)

Figure S8: linear scale of $|\nabla X_V|$ with a) the point defect generation rate $K_0$ for $d = 13.44$ nm and b) the grain size $d$ for $K_0 = 10^{-3}$ dpa/s.

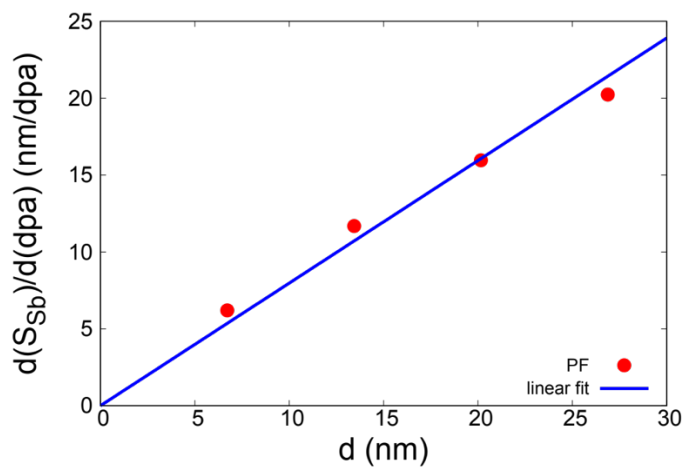

Figure S9: Linear scale of the initial rate of the total solute segregation length on the sink as a function of the grain size d.